\documentclass[preprints,article,accept,moreauthors,pdftex]{Definitions/mdpi} 

\firstpage{1} 
\pubvolume{xx}
\issuenum{1}
\articlenumber{5}
\pubyear{2020}
\copyrightyear{2020}
\history{Received: 28 March 2020; Accepted: 7 May 2020; Published: date}
\pdfoutput=1

\usepackage{amsmath,amssymb,txfonts}
\usepackage{graphicx,color}
\usepackage{subfigure}

\newcommand{\ergs}{{\rm \,erg\,s^{-1}}}

\newcommand{\mbh}{{M_{\rm BH}}}
\newcommand{\ledd}{{L_{\rm Edd}}}

\newcommand{\lx}{{L_{\rm X}}}
\newcommand{\lhx}{{L_{\rm HX}}}
\newcommand{\lr}{{L_{\rm R}}}

\newcommand{\pj}{{P_{\rm j}}}
\newcommand{\ldisk}{{L_{\rm disk}}}
\newcommand{\pnull}{{p_{\rm null}}}
\newcommand{\rx}{{R_{\rm X}}}

\Title{The Jet-Disk Coupling of Seyfert Galaxies from a Complete Hard X-ray Sample}

\Author{Xiang Liu $^{1,2,}$*, Ning Chang $^{1,3}$, Zhenhua Han $^{4}$, Xin Wang $^{1,3}$}

\AuthorNames{Xiang Liu, Ning Chang, Zhenhua Han, Xin Wang}

\address{%
$^{1}$ \quad Xinjiang Astronomical Observatory, Chinese Academy of Sciences, 150 Science 1-Street, Urumqi 830011, China; changning@xao.ac.cn (N.C.); wangxin2019@xao.ac.cn (X.W.)\\
$^{2}$ \quad Physics and Astronomy Department, Qiannan Normal University for Nationalities, Duyun 558000, China\\ % Please add the department/school/faculty/campus information, just like Affiliation 4.
$^{3}$ \quad School of Astronomy and Space Science, University of Chinese Academy of Sciences, Beijing 100049, China;\\
$^{4}$ \quad Physics and Electronic Engineering Department, Xinjiang Normal University, Urumqi 830000, China; hanzhh@xjnu.edu.cn}

\corres{Correspondence: liux@xao.ac.cn (X.L.)}

\abstract{We analyze the jet-disk coupling for different subsamples from a complete hard X-ray Seyfert sample to study the coupling indices and their relation to accretion rate. The results are: (1) the power-law coupling index ranges from nearly unity (linear correlation) for radio loud Seyferts to significantly less than unity for radio quiet ones. This decline trend of coupling index also holds from larger sources to compact ones; (2) the Seyferts with intermediate to high accretion rate (Eddington ratio $\lambda\sim$ 0.001 to 0.3) show a linear jet-disk coupling, but it shallows from near to super Eddington ($\lambda\sim$ 0.3 to 10), and the former is more radio loud than the latter; (3) the Seyfert 1s are slightly steeper than the Seyfert 2s, in the jet-disk correlation. In the linear coupling regime, the ratio of jet efficiency to radiative efficiency ($\eta/\varepsilon$) is nearly invariant, but in low accretion or super accretion regime, $\eta/\varepsilon$ varies with $\lambda$ in our model. We note that a radio-active cycle of accretion-dominated active galactic nuclei would be: from a weaker jet-disk coupling in $\lambda<0.001$ for low luminosity Seyferts, to a linear coupling in $0.001<\lambda<0.3$ for radio-loud luminous Seyferts and powerful radio galaxies/quasars, and to a weaker coupling in $0.3<\lambda<10$ ones.}

\begin{document}
%%%%%%%%%%%%%%%%%%%%%%%%%%%%%%%%%%%%%%%%%%

%%%%%%%%%%%%%%%%%%%%%%%%%%%%%%%%%%%%%%%%%%
%\setcounter{section}{-1} %% Remove this when starting to work on the template.
%\section{How to Use this Template}
%The template details the sections that can be used in a manuscript. Note that the order and names of article sections may differ from the requirements of the journal (e.g., the positioning of the Materials and Methods section). Please check the instructions for authors page of the journal to verify the correct order and names. For any questions, please contact the editorial office of the journal or support@mdpi.com. For LaTeX related questions please contact latex@mdpi.com.
%The order of the section titles is: Introduction, Materials and Methods, Results, Discussion, Conclusions for these journals: aerospace,algorithms,antibodies,antioxidants,atmosphere,axioms,biomedicines,carbon,crystals,designs,diagnostics,environments,fermentation,fluids,forests,fractalfract,informatics,information,inventions,jfmk,jrfm,lubricants,neonatalscreening,neuroglia,particles,pharmaceutics,polymers,processes,technologies,viruses,vision

\section{Introduction} \label{sec:intro}

Active galactic nuclei (AGNs) are the most powerful sources in the Universe and believed to be powered by their central massive black hole accretion. In the meantime, there are significant feedbacks from AGN to host galaxy. The accretion and feedback together have formed the co-evolution of the AGN and its host galaxy. In addition to the AGN winds and slow outflows to host galaxy, radio jet (collimated outflow) is a prominent feedback from AGN to host galaxy and to intergalactic medium, although the jetted phase of AGN is only about 10\% of its whole life from large AGN sample statistics \citep{condon13}. Thus, a big question is why AGN is radio silence (non-jetted) in most of its lifetime, but active in its whole lifetime in optical and probably in X-rays. Most AGNs are identified in optical field from local Seyfert galaxies to distant quasars.

For the complexity of AGN structure, their electromagnetic emissions are believed to be multi-components. The optical continuum (including UV and part of IR) and hard X-rays are thought to be mainly from the accretion disk and its corona. The bolometric disk luminosity is often scaled proportionally to the optical emission or hard X-rays from the disk, which is regulated by the mass accretion rate (or Eddington ratio) \citep{ho08,netzer19}. It is possible that an AGN evolves from low to high accretion rate, then reversely from high to low rate, for a lifetime (or episode) cycle of AGN. A jetted phase is present only in a part time ($\sim$10\%) of a life cycle of AGN, it is not clear which part of the accretion cycle (accretion rates) the jetted phase is most correlated.

We know that the accretion disk luminosity is proportional to the mass accretion rate in the form of $\ldisk=\varepsilon \dot{M}c^{2}$, with $\varepsilon$ the disk radiative efficiency, $c$ the speed of light. A jet power can be
\begin{equation}
\pj=1.3\times10^{38} (\eta /\varepsilon) [(\ldisk/\ledd)
M](\ergs)\propto(\eta /\varepsilon)\ldisk, \label{eq1}
\end{equation}
where $\eta$ is the jet efficiency depending on jet production mechanism, the Eddington
%please confirm all variables are correct and keep the format consistent in the whole paper.
 ratio $\lambda=\ldisk/\ledd$, with $\ledd$ the Eddington luminosity, $M$ the black hole (BH) mass in solar mass unit.

We can see that the jet-disk coupling is relative to the $\eta/\varepsilon$ in Equation (\ref{eq1}), which may be otherwise not constancy for different type of sources which being at different accretion rates \citep{xie16}. We absorb this unknown factor $\eta/\varepsilon$ in Equation (\ref{eq1}) into the power-law index $\mu$ in the form of
\begin{equation}
\pj=c_{1}\times (\eta/\varepsilon) \ldisk=c_{2}\times L^{\mu}_{\rm disk}=c_{3}\times L^{1+q}_{\rm disk}\propto(\lambda M)^{1+q}, \label{eq2}
\end{equation}
assuming $\eta/\varepsilon$ is also a function of the disk luminosity as $\eta/\varepsilon=function(\ldisk)=const\times L^{q}_{\rm disk}$, thus $\mu=1+q$, where $c_{1}$, $c_{2}$, and $c_{3}$ are constants. With this model, we can fit the relation of jet power and disk luminosity with $\mu$ and derive the $q$, in order to study the jet-disk coupling for different subsamples. These subsamples should come from a well-defined complete sample. {In the literature, there are discussions of the impact of magnetic flux \citep{sikora07} and the black hole spin \citep{Miller09,Unal20} on the jet-disk correlation, which may introduce some scatter in the correlation.}

In this paper, we reanalyze the complete hard-X-ray Seyfert sample by Panessa et al. \citep{panes15} in more detail, in order to investigate the jet-disk coupling for different subsamples. {In the Panessa sample, the 2--10 keV X-ray flux is from Malizia et al. \citep{Malizia09}, the radio data mainly come from the NRAO (National Radio Astronomy Observatory) VLA (Very Large Array) %please define NRAO and VLA if appropriate.
 Sky Survey (NVSS; Condon et al. \citep{condon98}) at 1.4 GHz, with additional complements being from the Sydney University Molonglo Sky Survey (SUMSS; Bock et al. \citep{Bock1999}) at 843 MHz, see Section \ref{sec:analysis} for more details. In order to avoid strong Doppler boosting effect (thus the luminosities measured are far away from their intrinsic values), sources like blazars are not included in this sample.}

%%%%%%%%%%%%%%%%%%%%%%%%%%%%%%%%%%%%%%%%%%
\section{Some Previous Statistics and Models} \label{sec:model}

There are several statistical findings that radio power almost linearly correlates with the emission line luminosity or accretion disk luminosity in powerful radio galaxies and quasars \citep{raw91,wil99,cao2004,van13}. These results are often explained by the accretion-jet model \citep{bland82,van13}. On the other hand, the statistical results of low luminosity AGNs (LLAGNs) show shallower correlations of jet-disk coupling \citep{ho2001,nagar05,ho09} with the power-law indices of 0.4--0.7 as summarized in \citep{su17}. This relative weak jet-disk coupling in the LLAGNs may be attributed either to an inefficient accretion \citep{yuan14} or to the BH spin-jet model \citep{bland77,tchek11,liu16,su17}.

There are also statistical results showing rather steeper (with power-law index of $\sim$1.4) radio-X-ray correlation for AGN \citep{dong14} and for BH X-ray binaries' outliers-track \citep{gal12,xie16}, which are possibly regulated by radiatively efficient accretion flow or disk-corona model \citep{hei03,fal04,qiao15}.

Panessa et al. \citep{panes15} presented a complete hard X-ray sample of relatively luminous Seyfert galaxies, and found significant correlation with slopes being consistent with those expected for radiatively efficient accreting systems. This complete sample consists of a couple of subsamples, which were not analyzed individually in \citep{panes15}.

%%%%%%%%%%%%%%%%%%%%%%%%%%%%%%%%%%%%%%%%%%
\section{The Subsample Analysis on Jet-Disk Coupling } \label{sec:analysis}

The Panessa et al. sample \cite{panes15} is a well-defined complete hard X-ray sample of relatively high luminosity AGN at $z < 0.36$, including Seyfert 1s and Seyfert 2s, with X-ray (2--10 keV and 20--100 keV) and radio data from NVSS at 1.4 GHz (or for some sources, the 843 MHz has been converted into 1.4 GHz flux density assuming a spectral index of -0.7, $S\propto \nu^{\alpha}$). The radio morphologies are classified as resolved (R), slightly resolved (S), and unresolved (U) (whose size smaller than one-half the restoring beam size at full width at half-maximum (FWHM = 45 arcsec), e.g., for a source at redshift of $\sim0.1$ typical in our study, the unresolved size is $\le23$ kpc) and the rms noise in the images are about 0.5--1 mJy/beam. Black hole mass is available for most of the sources. Therefore, we can estimate the radio jet power, bolometric disk luminosity ($20\lx$ (2--10 keV), \citep{panes15}), Eddington ratio, and X-ray radio loudness ($\rx=\lr/\lx$ (2--10 keV), with Log$\rx>-4.5$ for radio loud and Log$\rx<-4.5$ for radio quiet, as defined in \cite{tera03}). The mechanical radio jet power $\pj$ (rather than radio luminosity $L_{1.4}$ by \citep{panes15}) is used for including the work on environment by jet, expressed as $\pj\sim2.05\times10^{7}(L_{1.4})^{6/7}$ ($\ergs$) \citep{wil99,su17}, and the integrated flux density of source is adopted. The error of jet power is very small (not shown) in the log-log plot, considering an error of radio flux density of $<10\%$ for the VLA data. {Here, we assume that the model of jet power \cite{wil99} which derived mainly from massive BHs in FRI and FRII sources, can be used to the relatively lower BH masses of Seyfert galaxies in our sample.}

\begin{figure}[H]
	\subfigure{ \label{fig1a}
		\centering
		\includegraphics[width=7.5 cm]{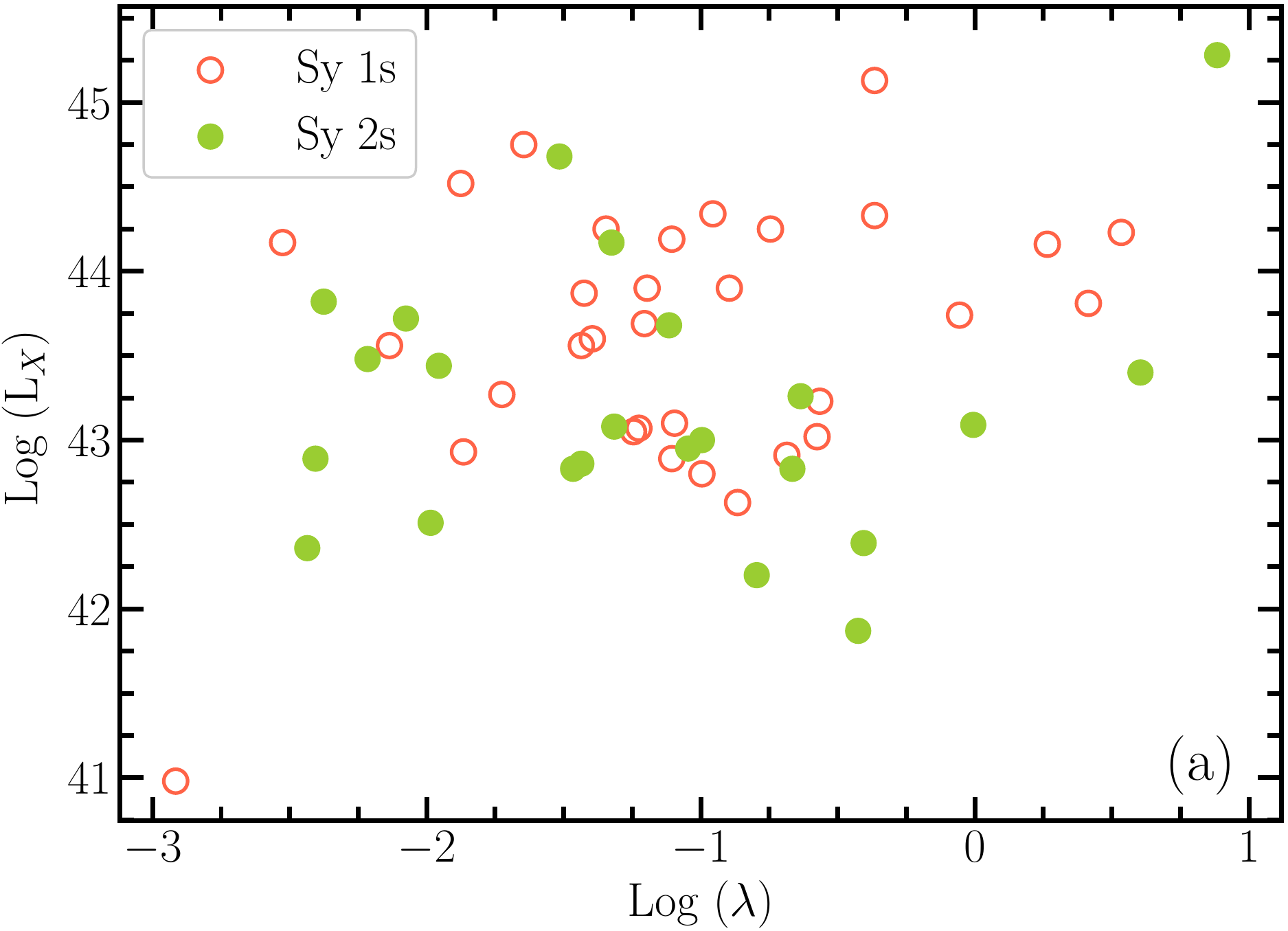}}
	\subfigure{ \label{fig1b}
		\centering
		\includegraphics[width=7.5 cm]{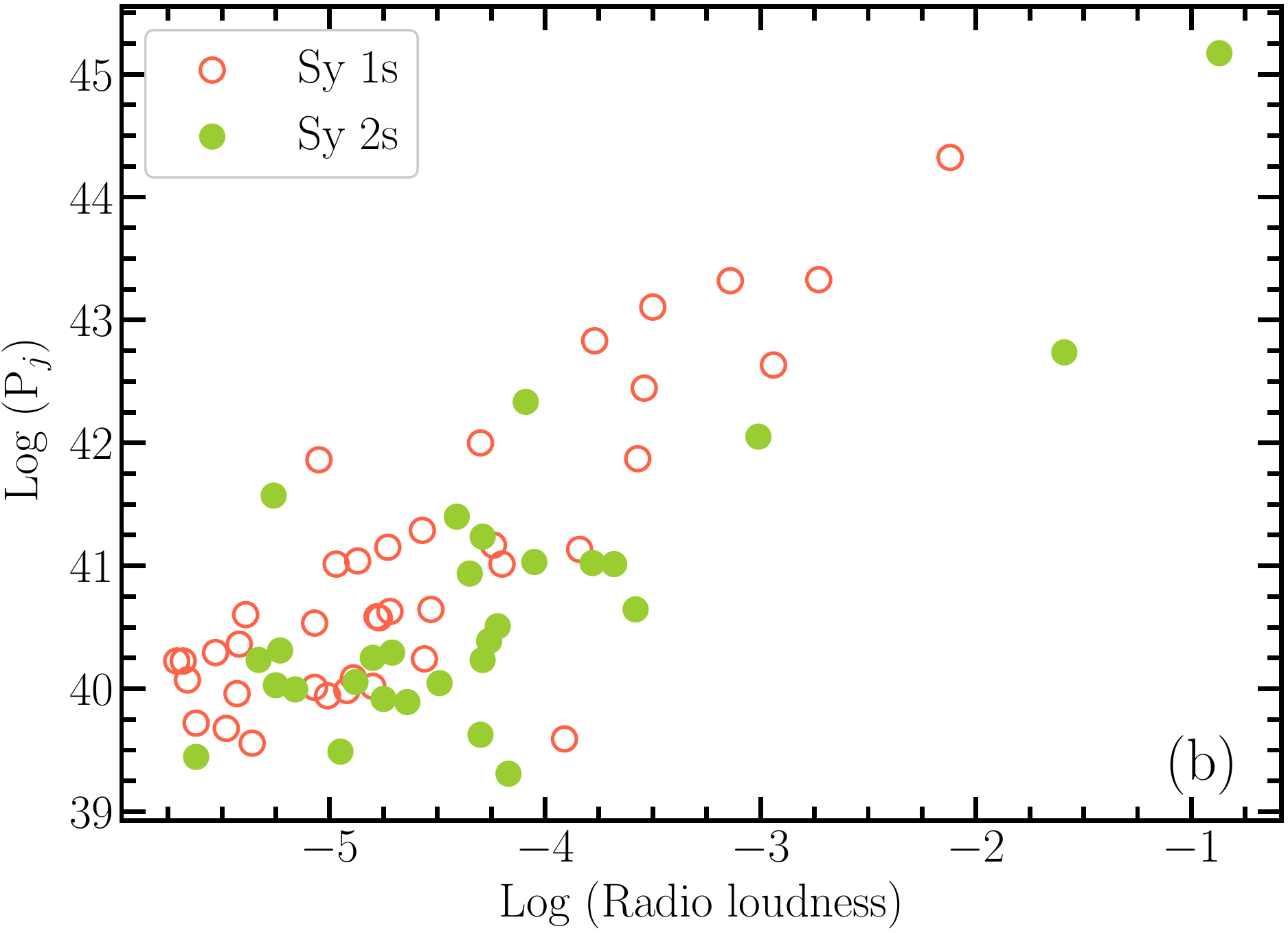}}
	\caption{\textbf{(a)} Log (X-ray luminosity in 2--10 keV) vs. Log (Eddington ratio $\lambda$) for Seyfert 1s (red) and 2s (green). \textbf{(b)} Log (jet power in $\ergs$) vs. Log (radio loudness) for Seyfert 1s (red) and 2s (green).}
	\label{fig1}
\end{figure}

From a total of 71 Seyfert galaxies with radio detections, we define two subgroups with 41 Seyfert 1s (including Seyfert 1 to Seyfert 1.5), and 30 Seyfert 2s (including Seyfert 1.9 to Seyfert 2) according to \cite{panes15}.
The BH mass in the two samples are comparable with the average of $10^{7.7}$, $10^{7.4}$, and median value of $10^{7.5}$, $10^{7.6}$ solar mass, for Seyfert 1s and 2s, respectively. The X-ray luminosity $\lx$ (2--10 keV) distributes roughly flat versus Eddington ratio as shown in Figure \ref{fig1a} for the Seyfert 1s and 2s. The radio jet power (at 1.4 GHz) is positively correlated with the radio loudness, and ranges similarly for the Seyfert 1s and 2s in Figure \ref{fig1b}.

\begin{figure}[H]
	\subfigure{ \label{fig2a}
		\centering
		\includegraphics[width=7.5 cm]{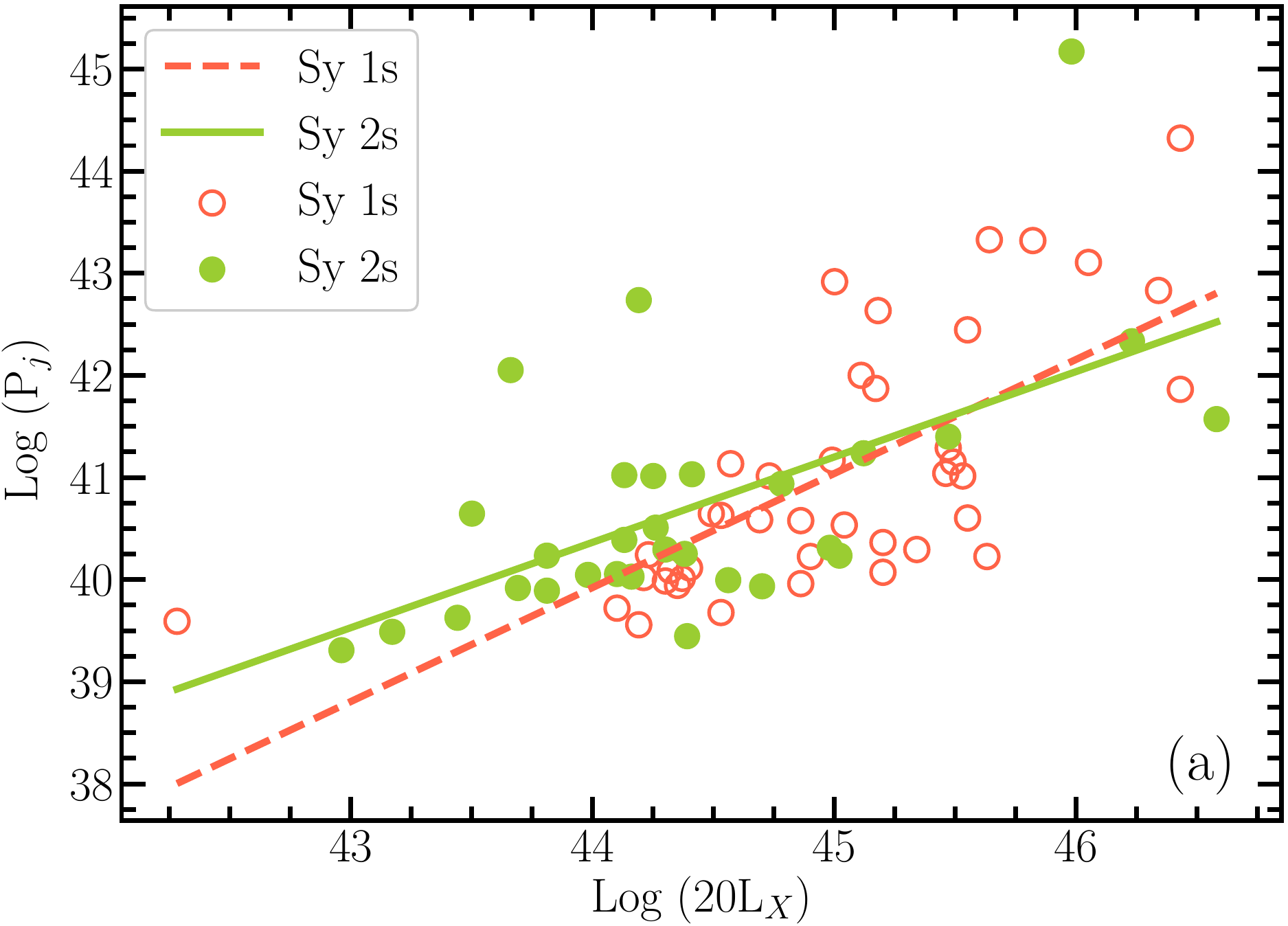}}
	\subfigure{ \label{fig2b}
		\centering
		\includegraphics[width=7.5 cm]{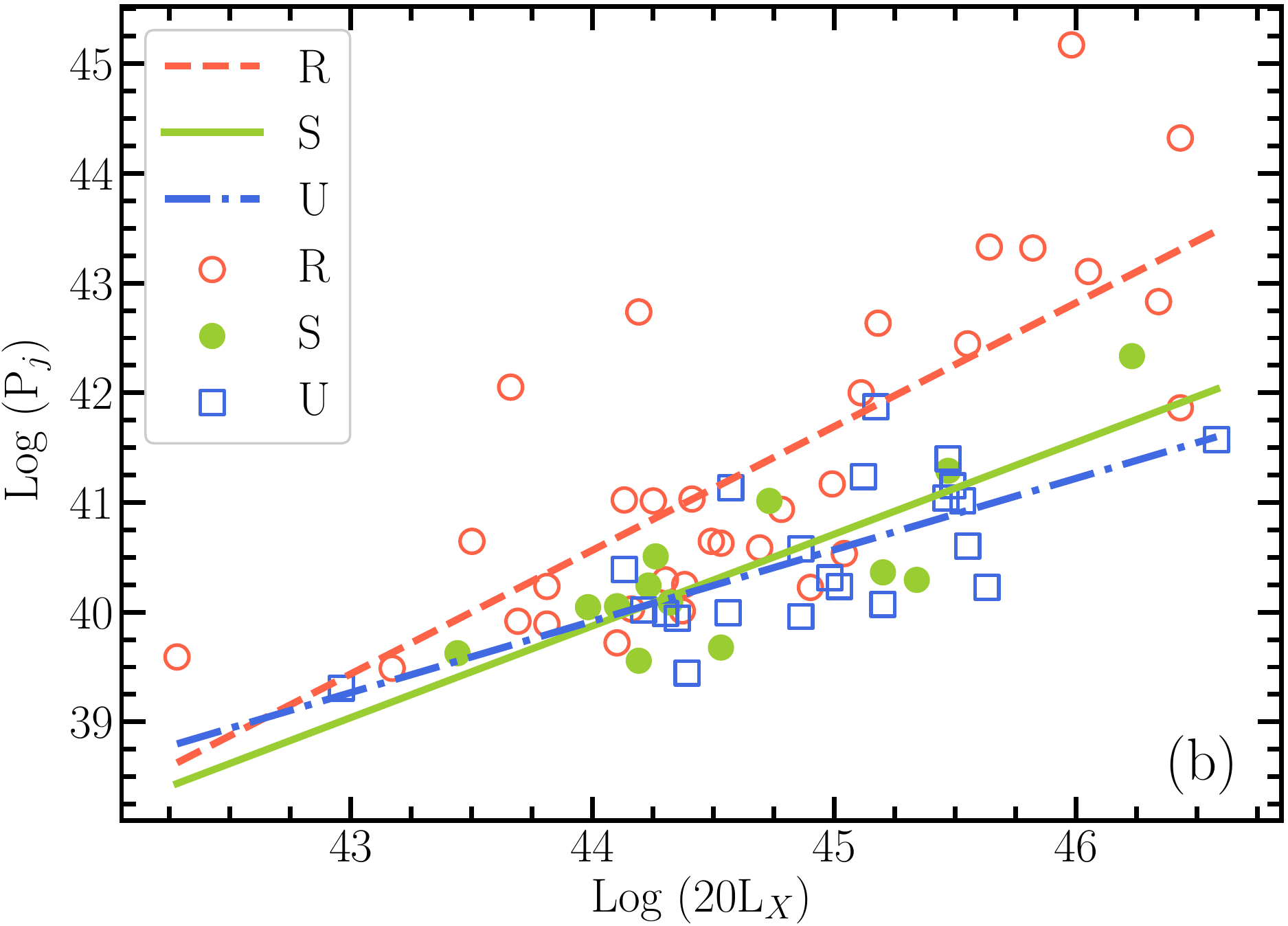}}
	
	\subfigure{ \label{fig2c}
		\centering
		\includegraphics[width=7.5 cm]{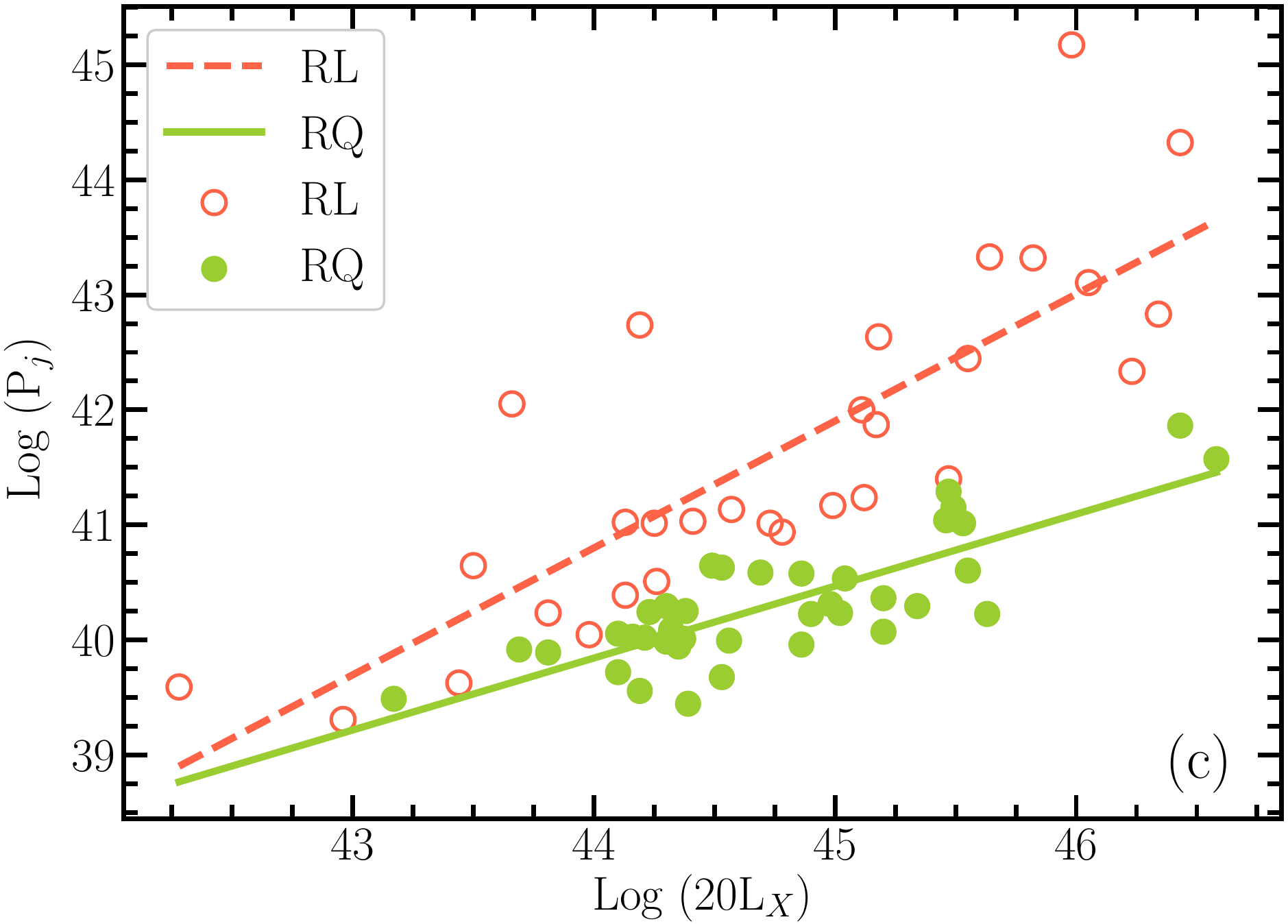}}
	\subfigure{ \label{fig2d}
		\centering
		\includegraphics[width=7.5 cm]{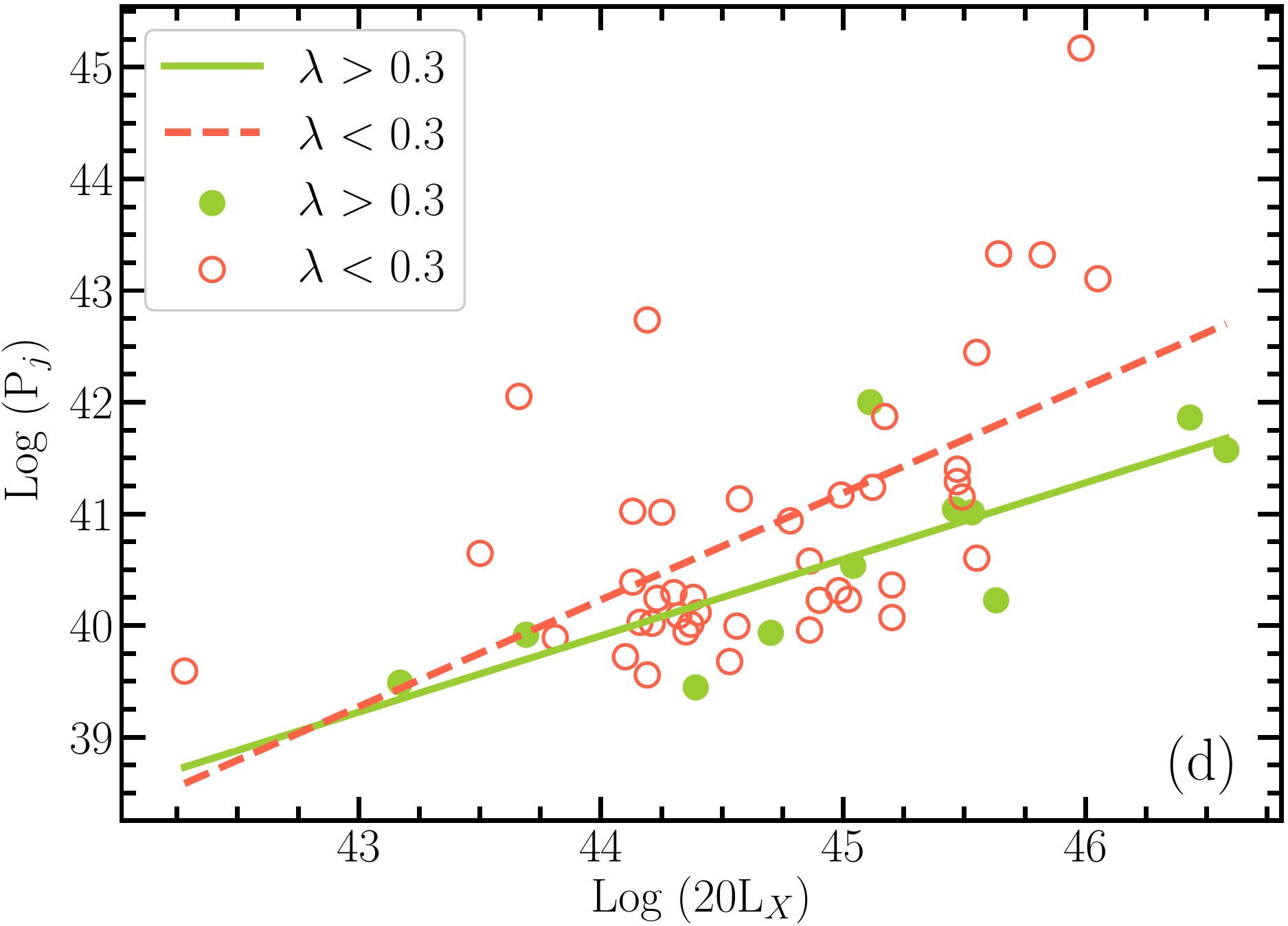}}		
	\caption{\textbf{(a)} Log (jet power in $\ergs$) vs. Log (disk luminosity in $\ergs$), with the best linear fit $y = (1.12\pm0.18)x - 9.22$, and $y = (0.84\pm0.21)x + 3.58$, for Seyfert 1s (red) and Seyfert 2s (green). \textbf{(b)} Log (jet power) vs. Log (disk luminosity), with the best linear fit $y = (1.13\pm0.17)x - 9.06$, $y = (0.84\pm0.18)x + 3.08$, and $y = (0.65\pm0.15)x + 11.27$ for resolved (red), slightly resolved (green), and unresolved radio sources (blue), respectively. \textbf{(c)} Log (jet power) vs. Log (disk luminosity), with the best linear fit $y = (1.10\pm0.15)x - 7.76$, and $y = (0.63\pm0.08)x + 12.32$ for radio loud (red) and radio quiet sources (green). \textbf{(d)} Log (jet power) vs. Log (disk luminosity), with the best linear fit $y = (0.68\pm0.18)x + 9.77$, and $y = (0.96\pm0.21)x - 1.89$ for sources with Eddington ratio $\lambda>0.3$ (green) and those with $\lambda<0.3$ (red).}
	\label{fig2}
\end{figure}

The coupling index between radio jet power and disk luminosity is shown in Figure \ref{fig2a}, for Seyfert 1s and 2s, with the steep index of $1.12\pm0.18$ in Seyfert 1s and $0.84\pm0.21$ in Seyfert 2s.

The fitted slopes are $1.13\pm0.17$, $0.84\pm0.18$, and $0.65\pm0.15$ for the resolved, slightly resolved, and unresolved sources, respectively in Figure \ref{fig2b}. 

The slopes for radio loud and radio quiet Seyferts are shown in Figure \ref{fig2c} with a nearly linear correlation ($1.10\pm0.15$) for radio loud and a shallower correlation ($0.63\pm0.08$) for radio quiet ones.

The slope of the subsample in near to super accretion rate (Eddington ratio $0.3<\lambda<10$) and that in moderate to high accretion rate ($0.001<\lambda<0.3$) is $0.68\pm0.18$ and $0.96\pm0.21$, respectively in Figure \ref{fig2d}. It is surprising that the jet-disk coupling index is lower at the very high accretion rates.

The slopes for larger BHs ($>10^{7.6}$ suns) and smaller BHs ($<10^{7.6}$ suns) are $0.75\pm0.34$ and $0.47\pm0.17$, respectively, with a large error of slope or a low significance of correlation in Table \ref{tab1}.

\begin{table}[H]
	
	\caption{The results of linear regression fits and correlation coefficients for Log$\pj$=$\mu$Log($20\lx$)+b, the columns are (1) subsample (Seyfert 1s, Seyfert 2s, resolved source by VLA, S: slightly resolved by VLA, U: unresolved by VLA, radio loud, radio quiet, Eddington ratio $>0.3$, Eddington ratio $<0.3$, BH mass $\mbh>10^{7.6}M_{\odot}$, BH mass $\mbh<10^{7.6}M_{\odot}$, respectively); (2) sample size; (3) median value of Log$\pj$ and Log($20\lx$); (4) power-law slope $\mu$ in $\pj=c_{1}\times (\eta/\varepsilon) \ldisk=c_{2}\times L^{\mu}_{\rm disk}=c_{3}\times L^{1+q}_{\rm disk}$, assuming $\eta/\varepsilon=const\times L^{q}_{\rm disk}$, where $\pj\sim2.05\times10^{7}(L_{1.4})^{6/7}$ ($\ergs$) and $\ldisk=20\lx$ (2--10 keV) ($\ergs$); (5)-(7) Pearson, Spearman, and Kendall correlation coefficient (and probability for rejecting the null hypothesis that there is no correlation), (8) $q=\mu-1$.}
	\centering
	%% \tablesize{} %% You can specify the fontsize here, e.g., \tablesize{\footnotesize}. If commented out \small will be used.
	
	\resizebox{\textwidth}{!}{
		\begin{tabular}{cccccccc}
			
			\toprule
			
			\textbf{Subsample}	& \textbf{Size}	& \textbf{Median($\pj,20\lx$)}	& \textbf{$\mu$}	& \textbf{Pearson($\pnull$)}	& \textbf{Spearman($\pnull$)}	& \textbf{Kendall($\pnull$)}	& \textbf{$q$}\\
			
			\textbf{(1)} & \textbf{(2)} & \textbf{(3)} & \textbf{(4)} & \textbf{(5)} & \textbf{(6)} & \textbf{(7)} & \textbf{(8)} \\
			
			\midrule
			
			Seyfert 1 & $41$ & $(40.6,\ 45.0)$ & $1.12\pm0.18$ & $0.70\ (2.9\times10^{-7}) $ & $0.75\ (1.5\times10^{-8})$ & $0.56\ (3.6\times10^{-7})$ & $0.12\pm0.18$ \\
			
			\hline
			
			Seyfert 2 & $30$ & $(40.3,\ 44.3)$ & $0.84\pm0.21$ & $0.59\ (5.5\times10^{-4})$ & $0.53\ (2.9\times10^{-3})$ & $0.40\ (2.1\times10^{-3})$ & $-0.16\pm0.21$ \\
			
			\hline
			
			Resolved & $33$ & $(40.9,\ 44.5)$ & $1.13\pm0.17$ & $0.76\ (2.4\times10{-7})$ & $0.73\ (1.5\times10^{-6})$ & $0.55\ (6.5\times10^{-6})$ & $0.13\pm0.17$ \\
			
			\hline
			
			S & $13$ & $(40.2,\ 44.3)$ & $0.84\pm0.18$ & $0.82\ (6.4\times10^{-4})$ & $0.79\ (1.5\times10^{-3})$ & $0.62\ (2.7\times10^{-3})$ & $-0.16\pm0.18$ \\
			
			\hline
			
			U & $22$ & $(40.4,\ 45.0)$ & $0.65\pm0.15$ & $0.69\ (3.5\times10^{-4})$ & $0.65\ (1.1\times10^{-3})$ & $0.43\ (4.8\times10^{-3})$ & $-0.35\pm0.15$  \\
			
			\hline
			
			R$_{loud}$ & $30$ & $(41.2,\ 44.8)$ & $1.10\pm0.15$ & $0.82\ (3.4\times10^{-8})$ & $0.84\ (5.8\times10^{-9})$ & $0.69\ (9.5\times10^{-8})$ & $0.10\pm0.15$ \\
			
			\hline
			
			R$_{quiet}$ & $38$ & $(40.2,\ 44.6)$ & $0.63\pm0.08$ & $0.81\ (7.7\times10^{-10})$ & $0.73\ (2.1\times10^{-7})$ & $0.53\ (3.7\times10^{-6})$ & $-0.37\pm0.07$ \\
			
			\hline
			
			$\lambda>0.3$ & $11$ & $(40.5,\ 45.1)$ & $0.68\pm0.18$ & $0.78\ (4.8\times10^{-3})$ & $0.76\ (7.3\times10^{-3})$ & $0.56\ (0.016)$ & $-0.32\pm0.18$ \\
			
			\hline
			
			$\lambda<0.3$ & $42$ & $(40.4,\ 44.6)$ & $0.96\pm0.21$ & $0.58\ (5.5\times10^{-5})$ & $0.54\ (2.0\times10^{-4})$ & $0.41\ (1.4\times10^{-4})$ & $-0.04\pm0.21$ \\
			
			\hline
			
			$\mbh>10^{7.6}$ & $29$ & $(41.2,\ 45.2)$ & $0.75\pm0.34$ & $0.42\ (0.025)$ & $0.47\ (9.7\times10^{-3})$ & $0.34\ (9.1\times10^{-3})$ & $-0.25\pm0.34$ \\
			
			\hline
			
			$\mbh<10^{7.6}$ & $24$ & $(40.0,\ 44.3)$ & $0.47\pm0.17$ & $0.52\ (9.2\times10^{-3})$ & $0.35\ (0.091)$ & $0.22\ (0.14)$ & $-0.53\pm0.17$ \\
			
			\bottomrule
			
	\end{tabular}}
	\label{tab1}
\end{table}

\begin{figure}[H]
	\subfigure{ \label{fig3a}
		\centering
		\includegraphics[width=7.5 cm]{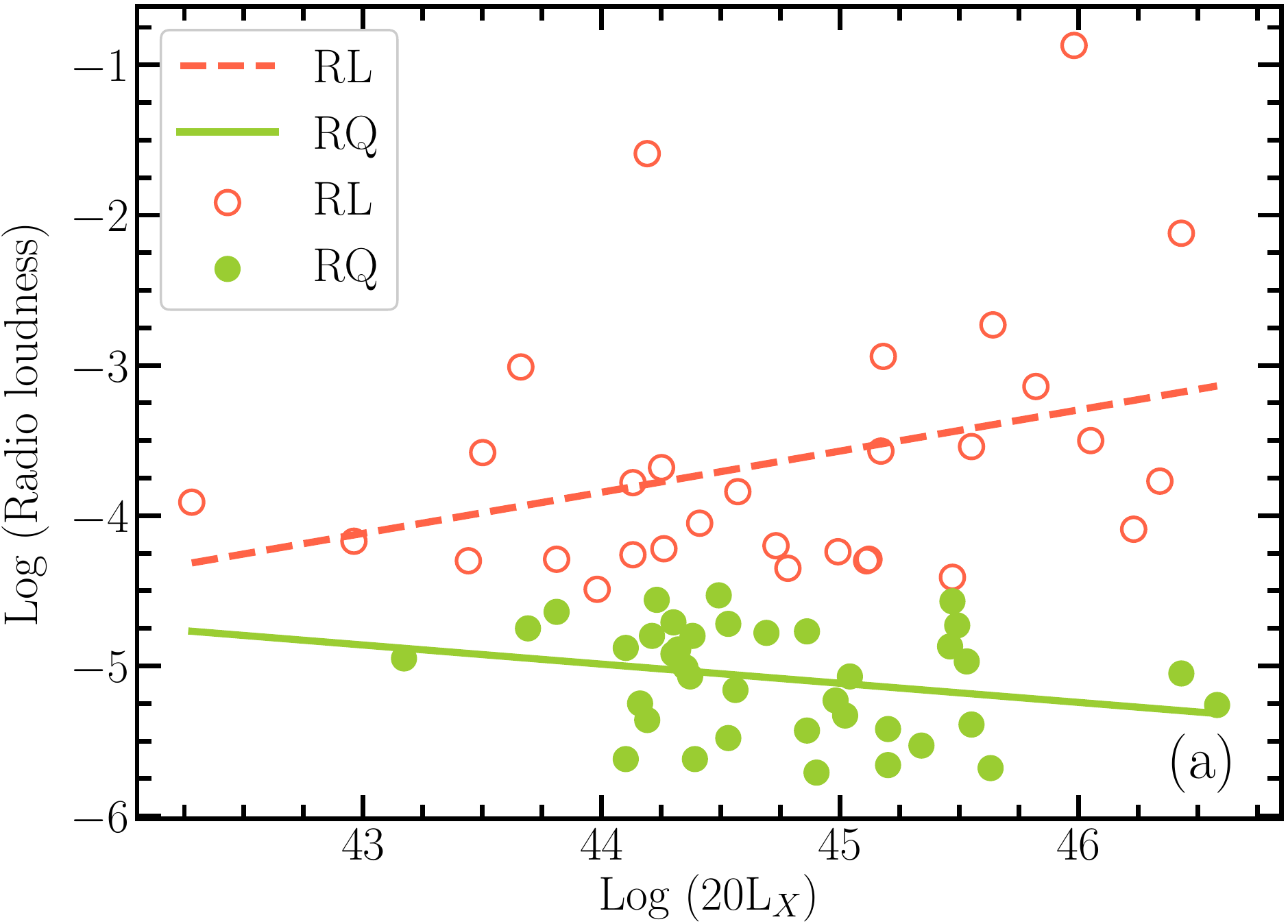}}
	\subfigure{ \label{fig3b}
		\centering
		\includegraphics[width=7.5 cm]{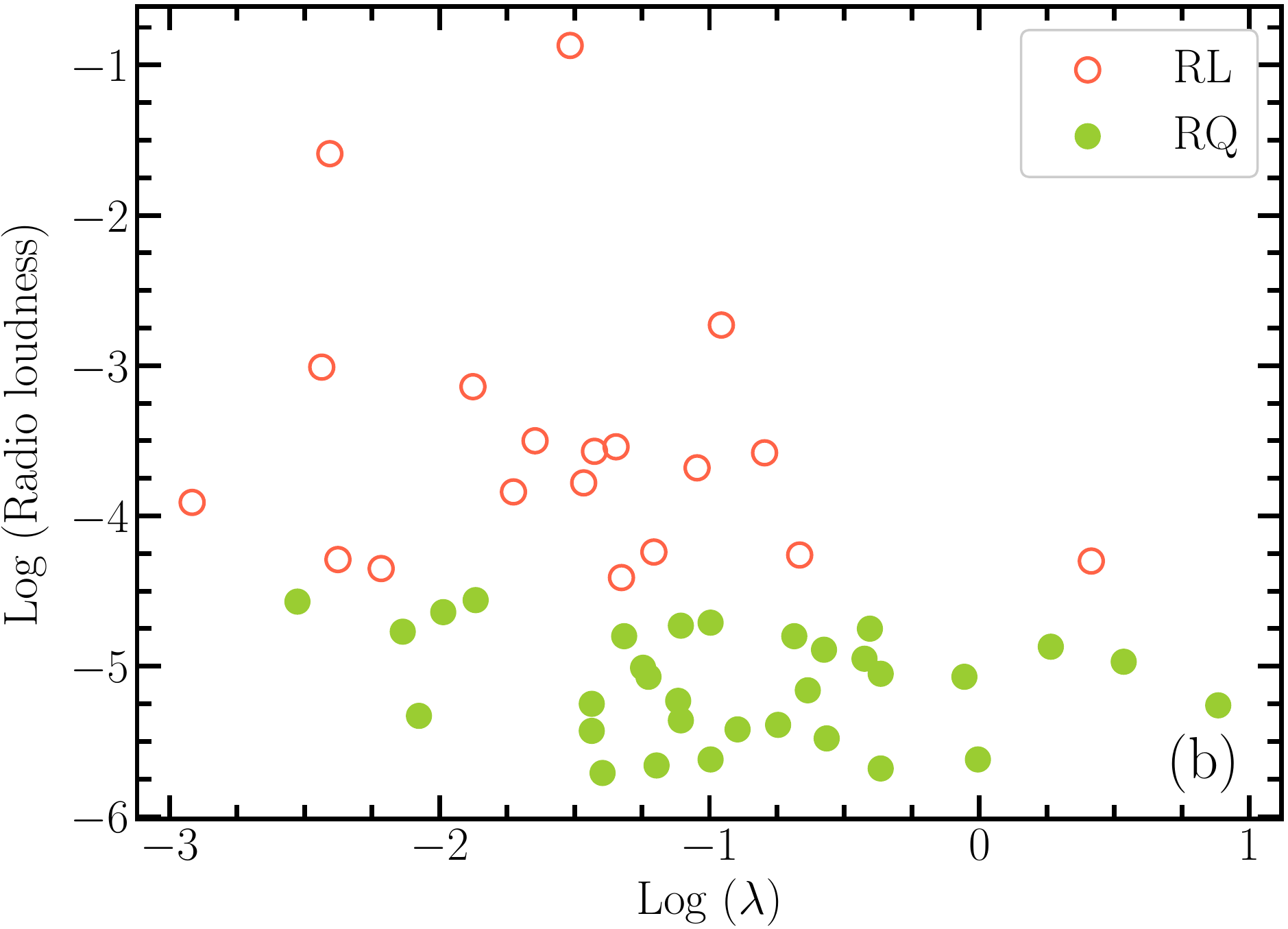}}
	\caption{\textbf{(a)} Log (radio loudness) vs. Log (disk luminosity in $\ergs$), with the best linear fit $y = (0.27\pm0.15)x - 15.89$, and $y = (-0.13\pm0.08)x +0.61$, for radio loud (red) and radio quiet sources (green). \textbf{(b)} Log (radio loudness) vs. Log (Eddington ratio) for radio loud (red) and radio quiet sources (green).}
	\label{fig3}
\end{figure}

The radio loudness versus the disk luminosity shows a bi-model alike distribution, the radio loud ones (RLs) show slightly positive correlation, and the radio quiet ones (RQs) show a slightly negative correlation in Figure \ref{fig3a}. These positive and negative correlations (with the power-law index of $\rho$) can be explained with the steeper index $\mu=1.1$ of jet-disk coupling in the RLs and the shallower $\mu=0.63$ in the RQs (Figure \ref{fig2c}), for $\pj\propto L_{1.4}^{6/7}$ and $\pj\propto (\ldisk)^{\mu}=(20\lx)^{\mu}$:

\begin{equation}
R_{loudness}=L_{1.4}/\lx\propto L^{(7/6)\mu-1}_{\rm X}=L^{\rho}_{\rm X} \propto L^{\rho}_{\rm disk},
\end{equation}
where $\rho=(7/6)\mu-1$.

For the whole sample, the radio loudness shows a decrease trend from Eddington ratio of 0.001 to 0.3 (or -3 to -0.5 in log scale), and has no significant correlation with the accretion rate in $0.3<\lambda<10$, in Figure \ref{fig3b}. {Implications of the Figure \ref{fig3b} are discussed in Section \ref{sec:discuss}.}

The results for the slope of jet-disk coupling, correlation coefficient (with probability) of the subsamples are summarized in Table \ref{tab1}. We will discuss the results further as following.

\section{Discussion} \label{sec:discuss}

We use the $20\lx$ (2--10 keV) as approximate bolometric disk luminosity. The higher energy luminosity in 20--100 keV ($\lhx$ in short) is also available in the Panessa et al. sample \cite{panes15}, who plotted the radio luminosity $L_{1.4}$ versus either $\lx$ (2--10 keV) or $\lhx$ (20--100 keV), showing that the radio-X-ray correlation slope is steeper in $\lhx$ band than in $\lx$ band. This is caused by the non-linear correlation of $\lhx\propto L^{0.84 \pm 0.04}_{\rm X}$ as we fitted. The $\lx$ (2--10 keV) is widely used to estimate the bolometric disk luminosity \citep[e.g.,][]{mer03,ho09,panes15}, and the high energy loss (or break) in the 20--100 keV band would prevent the $\lhx$ as a good estimate for bolometric disk luminosity.

The hard X-ray selected sample is almost unbiased with respect to absorption compared with the optical estimate for bolometric disk luminosity, and the Seyferts are mostly located in the nearby Universe, reducing as much as possible the selection (e.g., distance) effects \citep{Malizia09}. The Seyfert 1s, 2s, and R, S, U sources are equally distributed in redshift \citep{panes15}, and K-corrections are ignored in our analysis since the redshifts for our Seyferts are relatively small ($\leq 0.36$), and the correlation results in Table \ref{tab1} should be not caused by the redshift or distance effects \citep{gupta20}.

We obtained various jet-disk coupling indices (from 0.47 to 1.13) for different subsamples from the complete hard X-ray Seyfert sample. 
The linear correlation of jet-disk coupling ($\mu\sim1$), e.g., in the radio loud Seyferts, leads to $q=\mu-1=0$, i.e., a constant ratio of
$\eta/\varepsilon\propto L^{q=0}_{\rm disk}=constant$. This implies both the jet efficiency $\eta$ and disk radiative efficiency $\varepsilon$ are invariant, or they vary proportionally in the radio loud Seyferts. The linear correlation can be explained in Equation (\ref{eq2}) with the invariant ratio $\eta/\varepsilon$, and which is often referred to the accretion-dominated jet, as also shown in FRII quasars \citep{van13} and thought to be regulated by radiatively efficient accretion flow or disk corona model \citep{hei03,dong14,qiao15}.

The Seyfert 1s are usually more face-on to us than the Seyfert 2s, however, the beaming effect may not cause a steeper index of jet-disk coupling but can cause the scatter of the coupling as analyzed in \cite{hei04}. There are slightly larger median values of $\pj$ and $20\lx$ in the Seyfert 1s than in Seyfert 2s (see column 3 in Table \ref{tab1}), implying probably a potential contribution of $\lx$ from the jet base in Seyfert 1s.
This effect would be not significant as noted in \cite{gupta20}.

The index of jet-disk coupling, e.g., in the radio quiet sources (but not radio silence), is shallower, with $\mu<1, q=\mu-1<0$, i.e., $\eta/\varepsilon\propto L^{q<0}_{\rm disk}\propto(\lambda M)^{q<0}$. This implies the ratio of jet efficiency $\eta$ and disk radiative efficiency $\varepsilon$ varies with disk luminosity (or accretion rate and BH mass) for the radio quiet sources.

There is a decline trend of radio loudness vs. Eddington ratio $\lambda$ from 0.001 to 0.3, and no significant correlation with $\lambda$ from 0.3 to 10 in Figure \ref{fig3b} in the Seyfert galaxies. The decline part in the intermediate to high accretion regime shows a linear jet-disk coupling in Figure \ref{fig2d}, whereas in the near to super Eddington regime there is a shallower index ($\mu=0.68\pm0.18$) in Figure \ref{fig2d} with a low significance of correlation in Table \ref{tab1}. The result implies that very high accretion (near to super Eddington) may quench the jet, as noted in \cite{greene06}. There are similar decline trends between radio loudness and accretion rates as also found in \citep{ho02,sikora07,ho08}.

Furthermore, we used $20\lx$ (2--10keV) as an approximation of the bolometric disk luminosity, we note that in Vasudevan and Fabian \citep{Vasu2007}, the bolometric disk luminosity is about $20\lx$ (2--10keV) in Eddington ratio less than 0.2, but has a sharp increase to $\sim 60\lx$ (2--10keV) in $\lambda > 0.2$. We have tried to test the jet-disk correlations by using the two bolometric corrections of disk luminosity in the two accretion regimes respectively, the correlation results in Table \ref{tab1} are almost not changed. A future study should consider the bolometric correction dependence on accretion rate with a well-fitted function.

On the other hand, the low luminosity AGNs (Eddington ratio around $\lambda\sim0.0001$ or less) show smaller indices (0.4-0.7) of jet-disk coupling \citep{su17}, that could be explained by radiatively inefficient accretion flows (RIAF, \citep{yuan14}) and/or by the BH spin-powered jet (because of the spin-jet power being weakly correlated with the accretion rate, \citep[]{liu16,su17}).

In addition, we propose that a radio-active cycle of accretion-dominated AGN would be: from a weak jet-disk coupling ($\mu<1, q<0$) in low Eddington ratio ($\lambda<0.001$) in cosmic LLAGNs (excluding local radio-loud LLAGNs which we called `pseudo' radio-loud ones in the sense of a distant universe), to a linear (or even steeper) correlation ($\mu\ge1, q\ge0$) in intermediate to high Eddington ratio ($0.001<\lambda<0.3$) for radio-loud luminous Seyferts and powerful radio galaxies/quasars \citep{van13}, and to a weak coupling ($\mu<1, q<0$) again in near to supper Eddington ($0.3<\lambda<10$) (with a quenched/frustrated jet and a weak dependence of the ratio $\eta/\varepsilon$ on accretion rate).

In this scenario, the most of accretion-dominated AGNs may live in their radio loud phase ($\sim$10\% of AGN lifetime) in the middle regime of accretion rate with an efficient (linear or steeper) jet-disk coupling, especially for luminous Seyfert galaxies in this paper and for FRII quasars in \cite{van13}. In this idea we do not consider local LLAGNs that are about 50\% radio loud \citep{ho2001,nagar05}, for that the fraction will be much smaller for a distant universe due to radio undetectablity of the local LLAGNs) and their jet might not be accretion dominated but BH-spin dominated \citep{su17}.

Moreover, another fundamental factor seems to be changes in the spectral energy distribution (SED), which is driven by changes in the type of accretion disk (RIAF, standard disk, superluminous disk) for different regimes of accretion rates \citep{ho08,xie19}, that may partly be related to the function of ratio $\eta/\varepsilon$ on accretion rate and BH mass, and so the coupling of mechanical (jet) vs. radiative (disk) power, e.g., in Equation (\ref{eq2}).

\section{Summary and Conclusions} \label{sec:summary}

We reanalyzed the jet-disk coupling for various subsamples of a complete hard X-ray Seyfert sample in order to study the coupling indices and their relation to accretion rate. The results are: i) the power-law coupling index ranges from nearly unity (linear correlation) for radio loud Seyferts to significantly less than unity for radio quiet ones. This decline trend of coupling index also holds from larger sources to compact ones; ii) the Seyferts with intermediate to high accretion rate (Eddington ratio $\lambda\sim$0.001 to 0.3) show a linear jet-disk correlation, and the coupling shallows from near to super Eddington ($\lambda\sim$0.3 to 10), and the former is more radio loud than the latter; iii) the Seyfert 1s have a slightly steeper jet-disk coupling than the Seyfert 2s.

{A theoretical implication of the results is that,} in the linear coupling regime, the ratio of jet efficiency to radiative efficiency ($\eta/\varepsilon$) is nearly invariant, whereas in the low accretion or super accretion regime, $\eta/\varepsilon$ varies with $\lambda$ in our model of Equation (\ref{eq2}).

A radio-active cycle of accretion AGN would be: from a weak jet-disk coupling in low Eddington ratio ($\lambda<0.001$) for LLAGNs, to a linear correlation in intermediate to high Eddington ratio ($0.001<\lambda<0.3$) for radio-loud luminous Seyferts and powerful radio galaxies/quasars, and to a weak coupling again in near to supper Eddington ($0.3<\lambda<10$). In this scenario, the most of accretion-dominated AGNs may live as a radio loud source in the middle regime of accretion rate with an efficient (linear) jet-disk coupling, especially for radio-loud luminous Seyferts and quasars.

\vspace{6pt} 

%%%%%%%%%%%%%%%%%%%%%%%%%%%%%%%%%%%%%%%%%%
%% optional
%\supplementary{The following are available online at \linksupplementary{s1}, Figure S1: title, Table S1: title, Video S1: title.}


%%%%%%%%%%%%%%%%%%%%%%%%%%%%%%%%%%%%%%%%%%
\authorcontributions{Conceptualization, writing—original draft, X.L.; plotting and editing, N.C.; and review and editing, Z.H., X.W.. All authors have read and agreed to the published version of the manuscript.}%newly added, please confirm.

%%%%%%%%%%%%%%%%%%%%%%%%%%%%%%%%%%%%%%%%%%
\funding{This work is supported by the National Key R\&D Program of China under grant number 2018YFA0404602, and the Key Laboratory of Radio Astronomy, Chinese Academy of Sciences.}%please confirm the funding information. changes are not allowed after proofreading.

%%%%%%%%%%%%%%%%%%%%%%%%%%%%%%%%%%%%%%%%%%
\acknowledgments{We thank Luis C. Ho and Fu-Guo Xie for usefull comments on the manuscript.}

%%%%%%%%%%%%%%%%%%%%%%%%%%%%%%%%%%%%%%%%%%
\conflictsofinterest{The authors declare no conflict of interest.} 

%%%%%%%%%%%%%%%%%%%%%%%%%%%%%%%%%%%%%%%%%%
%% optional
%\abbreviations{The following abbreviations are used in this manuscript:\\
%
%\noindent 
%\begin{tabular}{@{}ll}
%MDPI & Multidisciplinary Digital Publishing Institute\\
%DOAJ & Directory of open access journals\\
%TLA & Three letter acronym\\
%LD & linear dichroism
%\end{tabular}}

%%%%%%%%%%%%%%%%%%%%%%%%%%%%%%%%%%%%%%%%%%
%% optional
%\appendixtitles{no} % Leave argument "no" if all appendix headings stay EMPTY (then no dot is printed after "Appendix A"). If the appendix sections contain a heading then change the argument to "yes".
%\appendix
%\section{}
%\unskip
%\subsection{}
%The appendix is an optional section that can contain details and data supplemental to the main text. For example, explanations of experimental details that would disrupt the flow of the main text, but nonetheless remain crucial to understanding and reproducing the research shown; figures of replicates for experiments of which representative data is shown in the main text can be added here if brief, or as Supplementary data. Mathematical proofs of results not central to the paper can be added as an appendix.
%
%\section{}
%All appendix sections must be cited in the main text. In the appendixes, Figures, Tables, etc. should be labeled starting with `A', e.g., Figure A1, Figure A2, etc. 

%%%%%%%%%%%%%%%%%%%%%%%%%%%%%%%%%%%%%%%%%%
\reftitle{References}

\end{document}